\documentclass[review]{elsarticle}
\usepackage{multirow} 
\usepackage{subfigure}
\usepackage{hyperref,color}
\makeatletter
\def\@author#1{\g@addto@macro\elsauthors{\normalsize%
    \def\baselinestretch{1}%
    \upshape\authorsep#1\unskip\textsuperscript{%
      \ifx\@fnmark\@empty\else\unskip\sep\@fnmark\let\sep=,\fi
      \ifx\@corref\@empty\else\unskip\sep\@corref\let\sep=,\fi
      }%
    \def\authorsep{\unskip,\space}%
    \global\let\@fnmark\@empty
    \global\let\@corref\@empty  
    \global\let\sep\@empty}%
    \@eadauthor={#1}
}
\makeatother

\journal{Journal of Information Processing and Management}

\begin{document}

\begin{frontmatter}

\title{Graph Neural News Recommendation with Long-term and Short-term Interest Modeling}


\author[BUPT]{Linmei Hu}
\ead{hulinmei@bupt.edu.cn}

\author[BUPT]{Chen Li}
\cortext[co]{Corresponding author}
\ead{leechen@bupt.edu.cn}

\author[BUPT]{Chuan Shi\corref{co}}
\ead{shichuan@bupt.edu.cn}

\author[BUPT]{Cheng Yang}
\ead{albertyang33@gmail.com}

\author[Ali]{Chao Shao}
\ead{shaochao.sc@alibaba-inc.com}

\address[BUPT]{Beijing University of Posts and Telecommunications, Beijing, China}
\address[Ali]{Alibaba Group, Hangzhou, China}


\begin{abstract}
With the information explosion of news articles, personalized news recommendation has become important for users to quickly find news that they are interested in. Existing methods on news recommendation mainly include collaborative filtering methods which rely on direct user-item interactions and content based methods  which characterize the content of user reading history. Although these methods have achieved good performances, they still suffer from data sparse problem, since most of them  fail to extensively exploit high-order structure information 
in news recommendation systems.  In this paper, we  propose  to build a heterogeneous graph  to explicitly model the interactions among users, news and latent topics. The incorporated topic information would help indicate a user's interest and alleviate the sparsity of user-item interactions. Then we take advantage of graph neural networks to learn  user and news representations that encode high-order structure information by  propagating embeddings over the graph.  The learned user embeddings with complete historic user clicks capture the users' long-term interests. We also consider a user's short-term interest using the recent  reading history with an attention based LSTM model. Experimental results on real-world datasets  show that our proposed model significantly outperforms  state-of-the-art methods  on news recommendation. 

\end{abstract}


\end{frontmatter}

\section{Introduction}
As the amount of  online news platforms such as Yahoo! news\footnote{https://news.yahoo.com/} and Google news\footnote{https://news.google.com/} increases, users are overwhelmed with a large volume of news from the worldwide covering various topics. To alleviate the information overloading, it is critical to help users target their reading interests and make personalized recommendations \cite{bansal2015content,li2010contextual,liu2010personalized,phelan2009using}. Therefore, news recommender systems that automatically recommend a small set of news articles  for satisfying user’s preferences, have growingly attracted attentions in both industry and academic \cite{das2007google,wang2017dynamic,DKN}.

There is a wide variety of typical methods to make personalized news recommendations, including collaborative filtering (CF)  methods \cite{das2007google,wang2011collaborative} and content based methods \cite{ijntema2010ontology,DKN,DAN,huang2013learning}. CF methods based on IDs always suffer from the cold start problem since out-of-date news are substituted by newer ones frequently. While content based methods completely ignore the collaborative signal. Hybrid methods combing CF and content for news recommendation have been proposed to address the problems \cite{de2012chatter,li2011scene}.
However,  all these methods  still  suffer  from  the  data  sparsity problem, since they fail to extensively exploit high-order structure information (e.g., the $u_1-d_1-u_2$ relationship indicates the behavior similarity between the users $u_1$ and $u_2$).  In addition, most of them ignore the latent topic information which would help indicate a user’s interest and alleviate the sparse user-item interactions. The intuition is that news items with few user clicks can  aggregate  more information with the bridge of topics. What's more, the existing methods on news recommendation do not consider the user's long-term and short-term interests. 
A user usually has relatively stable long-term interests and may also be temporally attracted to certain things, i.e., short-term interests, which should be considered in news recommendation. 



To address the above issues, in this paper, we propose a novel \textbf{\emph{G}}raph Neural \textbf{\emph{News}} \textbf{\emph{Rec}}ommendation model (\textbf{\emph{GNewsRec}}) with long-term and short-term user interest modeling. We first construct a heterogeneous user-news-topic graph as shown in Figure \ref{fig 1} to explicitly model the interactions among users, news and topics with complete historic user clicks. The topic information can help better reflect a user's interest and alleviate the sparsity issue of user-item interactions. To encode the high-order relationships among users, news items and topics, we take advantage of graph neural networks (GNN) to learn user and news representations  by  propagating embeddings over the graph.  The learned user embeddings with complete historic user clicks are supposed to encode a user's long-term interest. We also model a user's short-term interest using recent user reading history with an attention based LSTM \cite{LSTM,liu2018cross} model. We combine both long-term and short-term interests for user modeling, which are then compared to the candidate news representation for prediction. Experimental results on  real-world datasets show that our model significantly outperforms state-of-the-art methods on  news recommendation.

Our main contributions can be summarized as follows:
\begin{itemize}
\item[1)] In this work, we propose a novel graph neural news recommendation model GNewsRec  with long-term and short-term user interest modeling.
\item[2)] Our model constructs a heterogeneous user-news-topic graph to model user-item interactions, which alleviates the sparsity of user-item interactions. Then it applies  graph  convolutional networks to learn user and news embeddings with high-order information encoded by propagating embeddings over the graph.
\item[3)] Experimental results on real-world datasets demonstrate that our proposed model significantly outperforms sate-of-the-art methods on news recommendation.
\end{itemize}

\section{Related Work}
Personalized news recommendation has been widely studied with the information overloading of news articles. A variety of methods have been proposed including collaborative filtering (CF) based methods \cite{das2007google, marlin2004multiple, rendle2010factorization, li2015deep, wu2016collaborative, DMF}, and content based methods \cite{ijntema2010ontology, kompan2010content, huang2013learning}. 

CF methods assume  that behaviorally similar users would exhibit similar preference on items.  They  parameterize users and items for reconstructing historical interactions, and predict user preferences based on the parameters \cite{cao2019unifying, xiangnanhe, he2017neural}. For example, matrix factorization (MF) directly embeds user/item ID as vectors and models user-item interaction with inner product. DMF \cite{DMF} is a deep matrix factorization model which uses multiple nonlinear layers to process both explicit ratings and implicit feedback of users and news. However, most existing CF-based methods build the user and item embeddings with the descriptive features only (e.g., ID and attributes), without considering the higher-order information within the user-item interaction graph. Wang et al. \cite{xiangnanhe} proposed a neural graph CF method which exploits the user-item graph structure by propagating embeddings over the graph.
While CF methods still suffer from the cold-start problem since news items are substituted frequently.

To address this issue, content-based or hybrid methods have been proposed \cite{DeepWide, DeepFM, huang2013learning, zhang2018deep, DKN, DAN}. Content based methods consider the actual content or attributes of the items for making recommendations. For example,  DeepWide \cite{DeepWide} combines the linear model(Wide) and feed-forward neural network(Deep) to model feature interactions simultaneously. DeepFM \cite{DeepFM} integrates a component of factorization machines and a component of deep neural networks to model low-level and high-level feature interactions, respectively. 
DKN \cite{DKN} proposed a  news recommendation framework that fuse semantic-level and knowledge-level representations of news by a multi-channel CNN, and uses an attention module to dynamically calculate a user’s aggregated historical representation. DeepJONN \cite{zhang2018deep} is a session-based model which uses a tensor based CNN to model the session representation and an RNN to capture sequential patterns in streams of clicks
and associated features.  DAN \cite{DAN} improves DKN by designing an attention-based RNN to capture sequential information of clicked news, which achieves the state-of-the-art performance on news recommendation.  
Hybrid methods \cite{li2011scene,liu2010personalized, cantador2011enhanced} such as SCENE \cite{li2011scene}   usually combine several different recommender algorithms to recommend items.

Different from the above works, in this paper, we propose a novel graph neural news recommendation model with long-term and short-term interest modeling. It is  a hybrid method utilizing both user-item interactions and the contents of news articles. Our method extensively exploits the high-order structure information between users and items by constructing a heterogeneous graph and applying graph convolutional networks to propagate the embeddings.

\section{Problem Formulation}
The news recommendation problem in our paper can be illustrated as follows. We have the click histories for $K$ users $U = \{u_1, u_2,\cdot\cdot\cdot,u_K \}$ over $M$ news items $I = \{d_1, d_2,\cdot\cdot\cdot,d_M\}$. The user-item interaction matrix $Y \in R^{K \times M}$ is defined according to users' implicit feedback, where $y_{u,d} = 1$ indicate the user $u$ clicked the news $d$, otherwise $y_{u,d} = 0$. Additionally, from the click history with timestamps, we can obtain the recent click sequence  $s_u=\{d_{u,1}, d_{u,2},\cdot\cdot\cdot, d_{u,n}\}$ for  a specific user $u$, where $d_{u,j} \in I$ is the $j$-th news the user $u$ clicked. 

Given the user-item interaction matrix $Y$ as well as the users' recent click  sequences $S$, we aim to predict whether a user $u$ has potential interest in a news item $d$  which he/she has not seen before.
This paper considers the title and profile (a given set of entities $E$ and their entity types $C$  from the news page content) of news as features. Each news title $T$ contains a sequence of words $T = \{w_1, w_2, \cdot\cdot\cdot, w_{m}\}$.  The profile contains a sequence of entities  $E = \{e_1, e_2, \cdot\cdot\cdot, e_{n}\}$ as well as its  type set $C = \{c_1, c_2, \cdot\cdot\cdot, c_{n}\}$, where $c_j$ is the type of the $j$-th entity $e_j$.
\section{The Proposed Method}

\begin{figure}[t]
\centering
\includegraphics[width=0.8\textwidth]{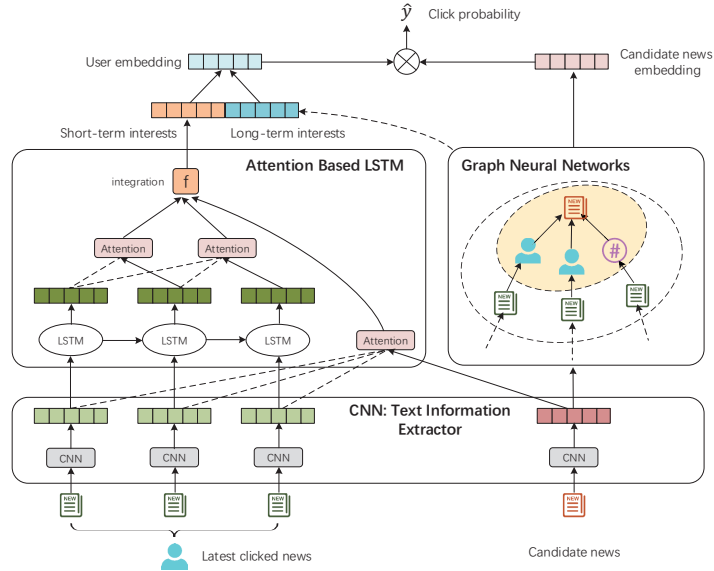}
\caption{The framework of GNewsRec.}
\label{fig 2}
\end{figure}

In this section, we present our graph neural news recommendation model GNewsRec with long-term and short-term interest modeling. Our model takes full advantage of the high-order structure information between users and news items by first constructing a heterogeneous graph modeling the interactions and then applying GNN to propagate the embeddings.  As illustrated in Figure \ref{fig 2},  GNewsRec contains three main parts: CNN for text information extraction, GNN for long-term user interest modeling and news modeling, and attention based LSTM model for short-term user interest modeling. The first part extracts the news feature from the news title and profile through CNN. The second part constructs a heterogeneous user-news-topic graph with complete historic user clicks and applies GNN to encode high-order structure information for recommendation. The incorporated latent topic information can alleviate the user-item sparsity since news items with few user clicks can aggregate more information  with the bridge of topics.  The learned user embeddings with complete historic user clicks are supposed to encode the relatively stable long-term user interest. We also model the user's short-term interest with recent reading history through an attention based LSTM in the third part. Finally, we combine a user's long-term  and short-term interests for user representation, which is then compared to candidate news representation for recommendation.
We will detail the three parts as follows.



\subsection{Text Information Extractor}
We use two parallel CNNs as the news text information extractor, which respectively take the title and profile of news as inputs and learn the title-level and profile-level representations of news. The concatenation of such two representations is regarded as the final text feature representation of news. 

Specifically, we represent the title as $\mathbf{T}=[\mathbf{w}_1,\cdot\cdot\cdot,\mathbf{w}_m]^T$ and  the profile as $\mathbf{P} = [\mathbf{e}_1, f(\mathbf{c}_1), \mathbf{e}_2, f(\mathbf{c}_2),\cdot\cdot\cdot ,\mathbf{e}_n, f(\mathbf{c}_n)]^T$, 
where $\mathbf{P} \in  R^{2n \times k_1}$ and $k_1$ is the dimension of entity embedding. $f(\mathbf{c}) = \mathbf{W}_c \  \mathbf{c}$ is the transformation function. $\mathbf{W}_c \in R^{k_1 \times k_2}$ ($k_2$ is the dimension of entity type embedding) is the trainable transformation matrix.


The title $\mathbf{T}$ and profile   $\mathbf{P}$ are respectively fed into two parallel CNNs that have separate weight parameters.  Hence  we separately obtain their feature representations as $\mathbf{\widetilde T}$ and $\mathbf{\widetilde P}$ through two parallel CNNs. Finally we concatenate $\mathbf{\widetilde T}$ and $\mathbf{\widetilde P}$ as the final news text feature representation:
\begin{equation}
\mathbf{d} = f_c([\mathbf{\widetilde T};\mathbf{\widetilde P}]),
\end{equation}
where $\mathbf{d} \in R^D$ and $f_c$ is a densely connected layer. 
\subsection{Long-term User Interest Modeling and News Modeling}
To model long-term user interest and news, we first construct a heterogeneous user-news-topic graph with users' complete historic clicks. The incorporated topic information can help better indicate a user's interest and alleviate the sparsity of user-item interactions. Then we apply  graph convolutional  networks for learning embeddings of users and news items, which encodes the high-order information between users and items through propagating embeddings over the graph.

\subsubsection{Heterogeneous User-News-Topic Graph}
\begin{figure}[t]
\centering
\includegraphics[width=0.8\textwidth]{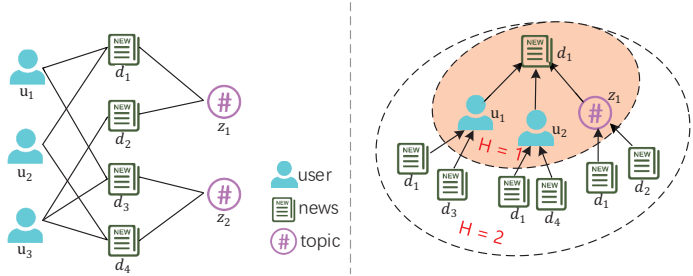}
\caption{Heterogeneous user-news-topic graph(left) and two-layers GNN(right).}
\label{fig 1}
\end{figure}
We incorporate the latent topic information in news articles to better indicate the user's interest and alleviate the user-item sparsity issue. Hence, we construct a heterogeneous undirected graph $G=(V,R)$ as illustrated in the left part of Figure \ref{fig 1}, where $V$ and $R$ are respectively the sets of nodes and edges. Our graph contains three types of nodes: users $U$, news items $I$ and topics $Z$. The topics $Z$ can be mined through the topic model LDA \cite{LDA}.

We build the user-item edges if the user $u$ clicked a news item $d$, i.e., $y_{u,d} =1$. For each news document $d$, we can obtain its topic distribution $\theta_d=\{\theta_{d,i}\}_{i=1,\cdot\cdot\cdot,\mathcal{K}}$, $\sum_{i=1}^{\mathcal{K}}\theta_i = 1$ through LDA. We build the connection of the news document $d$ and the topic $z$ with the largest probability. 

Note that for testing, we can infer the topics of new documents based on the estimated LDA model \cite{newman2008distributed}. In this way, the new documents that do not existed in the graph can be connected with the constructed graph and update their embeddings through graph convolution. Hence, the topic information can alleviate the cold start problem as well as the sparsity issue of user-item interactions.

\subsubsection{GNN for Heterogeneous User-News-Topic Graph}
With the constructed heterogeneous user-news-topic graph, we then apply GNN \cite{graphsage,KGCN,wang2019kgat} to capture high-order relationships between users and news by propagating the embeddings through it. 

In particular, consider the candidate pair of user $u$ and news $d$. We use $U(d)$ and $Z(d)$ \footnote{In this paper, we assume each news has only one topic, i.e., $|Z(d)|=1$} to respectively denote the set of users and topics directly connected to the news document $d$. In real applications, the size of $U(d)$ may vary significantly over all news documents. To keep the computational pattern of each batch fixed and more efficient, we uniformly sample a set of neighbors $S(d)$ with fixed size for each news $d$ instead of using its full neighbors, where the size $|S(d)| = L_u$.\footnote{$S(d)$ may contain duplicates if $|U(d)| < L_u$. If $U(d) = \emptyset$, then $S(d) = \emptyset$.}  


To characterize the topological proximity structure of news $d$, firstly, we compute the  linear average combination of all its sampled neighbors:
\begin{equation}
{\rm \mathbf{d}}_{\mathcal{N}} = \frac{1}{|S(d)|}\sum_{u \in S(d)} {\rm {\rm \mathbf{W}_u}\mathbf{u}} + \frac{1}{|Z(d)|}\sum_{z \in Z(d)} {\rm \mathbf{W}_z}{\rm \mathbf{z}},
\end{equation}
where $\mathbf{u} \in R^D$ and $\mathbf{z} \in R^D$ are the representations of the neighboring user and topic of news $d$. $\mathbf{u}$ and $\mathbf{z}$  are initialized randomly, while $\mathbf{d}$ are initialized with the text feature embeddings obtained from text information extractor (Sectioin 4.1). $\mathbf{W}_u \in R^{D \times D}$ and $\mathbf{W}_z \in R^{D \times D}$ are respectively the trainable transformation matrix for users and topics, which map them from the different spaces to the same space of news embeddings.

Then we update the candidate news embedding with the neighborhood representation $\mathbf{d}_{\mathcal{N}}$ by:
\begin{equation}
{\rm \mathbf{{\widetilde d}}} = \sigma({\rm \mathbf{W}}^{1}\cdot {\rm \mathbf{d}}_{\mathcal{N}} +  {\rm \mathbf{b}}^{1}),
\end{equation}
where $\sigma$ is the nonlinear function $ReLU$, and $\mathbf{W}^{1} \in R^{D \times D}$ and $\mathbf{b}^{1} \in R^D$ are  transformation weight and bias of the first layer of GNN, respectively.

This is a single layer GNN, where the final embedding of the candidate news is only dependent on its immediate neighbors. In order to capture high-order relationships between users and news, we  can extend the GNN from one layer to multiple layers,  propagating the embeddings in a broader and deeper way. As shown in Figure \ref{fig 1},  2-order news embeddings can be obtained as follows. We first get its 1-hop neighboring user embeddings $\mathbf{u}_l$ and topic embeddings $\mathbf{z}$ by aggregating their neighboring news embeddings.
Then we aggregate their embeddings $\mathbf{u}_l$ and $\mathbf{z}$ to get 2-order news embeddings $\mathbf{\widetilde d}$. Generally speaking, the $H$-order representation of an news is a mixture of initial representations of its neighbors up to $H$ hops away.

Through the GNN, we can get the final user and news embeddings $\mathbf{u}_l$ and $\mathbf{\widetilde d}$ with high-order information encoded. The user embeddings learned with complete user click history are supposed to capture the relatively stable long-term user interests. However, we argue that a user could be temporally attracted to certain things, namely, a user has short-term interest, which should also be considered in personalized news recommendation.

\subsection{Short-term User Interest Modeling}
In this subsection, we present how to model a user's  short-term interest using her recent click history through an attention based LSTM model. We  pay attention to not only the news contents but also  the sequential information.

\textbf{Attention over Contents}. Given a user $u$ with her latest $l$ clicked news  $\{\mathbf{d}_1, \mathbf{d}_2, ..., \mathbf{d}_l\}$\footnote{If the click history sequence  length is less than $l$, we pad it with zero embeddings.}, we use an attention mechanism to model the different impacts of the user's recent clicked news on the candidate news $d$:
 \begin{equation}
\mathbf{u}_j = tanh(\mathbf{W'}\mathbf{d}_j + \mathbf{b'}),
\end{equation}
\begin{equation}
{\rm \mathbf{u}} = tanh({\rm \mathbf{Wd + b}}),
\end{equation}
 \begin{equation}
\alpha_{j} = \frac{exp(\mathbf{v}^T( \mathbf{u}+\mathbf{u}_j))}{\sum_j exp( \mathbf{v}^T( \mathbf{u}+\mathbf{u}_j))},
\end{equation}
\begin{equation}
\mathbf{u}_c = \sum_j \alpha_{j} \mathbf{d}_j,
\end{equation}
where $\mathbf{u}_c$ is the user's current content-level interest embedding, $\alpha_{j}$ is the impact weight of clicked news $d_j (j=1,\cdot\cdot\cdot,l)$ on candidate news $d$,  $\mathbf{W}'$, $\mathbf{W}$ $\in R^{D \times D}$, $\mathbf{d}_j$, $\mathbf{b}_w$, $\mathbf{b}_t$, $\mathbf{v}^T$ $\in R^D$, $D$ is the dimension of news embedding.

\textbf{Attention over Sequential Information}. Besides applying attention mechanism to model user current content-level interest, we also take attention of the sequential information of the latest clicked news, thus we use an attention based LSTM \cite{LSTM} to capture the sequential features. 

As is shown in Figure \ref{fig 2}, LSTM takes user's clicked news embeddings as input, and output the user's sequential feature representation. Since each user's current click is affected by previous clicked news, the attention mechanism described above(for content-level interest modeling) is applied on each hidden state $\mathbf{h}_j$ and their previous hidden states $\mathbf{\{h}_1,\mathbf{h}_2,\cdot\cdot\cdot, \mathbf{h}_{j-1}\}$ ($\mathbf{h}_j=$ LSTM($\mathbf{h}_{(j-1)}$,$\mathbf{d}_j$)) of the LSTM to obtain richer sequential feature representation $\mathbf{s}_j,(j=1,\cdot\cdot\cdot,l)$ at different click times. These features $(\mathbf{s}_1,\cdot\cdot\cdot,\mathbf{s}_l)$ are integrated by a CNN to get the final sequential feature representation {$\mathbf{\widetilde s}$} of user's latest $l$ clicked history.

We feed the concatenation of current content-level interest embedding and the sequence-level embedding into a fully connected network and get the final user's short-term  interest embedding:
\begin{equation}
\mathbf{u}_s = \mathbf{W}_s[\mathbf{u}_c; \widetilde s],
\end{equation}
where $\mathbf{W}_s \in R^{D \times 2D}$.

\subsection{Prediction and Training}
Finally, the user embedding $\mathbf{u}$ is computed by taking linear transformation over the concatenation of the long-term and short-term embedding vectors:
\begin{equation}
\mathbf{u} = \mathbf{W}[\mathbf{u}_l; \mathbf{u}_s],
\end{equation}
where $\mathbf{W} \in R^{d \times 2d}$.

Then we compare the final user  embedding  $\mathbf{u}$ to the candidate news embedding
{$\mathbf{\widetilde d}$}, the probability of user $u$ clicking news $d$ is predicted
by a DNN:
\begin{equation}
\hat{y} = DNN(\mathbf{u},  \mathbf{\widetilde d}).
\end{equation}

To train our proposed model GNewsRec, we select positive samples from the existing observed clicked reading history  and equal amount of negative samples from unobserved reading. A training sample is denoted as $X = (u$, $x, y)$, where  $x$ is the candidate news to predict whether click or not. For each positive input sample, $y = 1$, otherwise $y = 0$. After our model, each input sample has a respective estimated probability $\hat{y} \in [0,1]$ of the user whether will click the candidate news $x$. We use the cross-entropy loss as our lost function:
\begin{equation}
\mathcal{L} = -\{\sum_{X \in \Delta^+}y\ log\ \hat{y} + \sum_{X \in \Delta^-}(1-y)\ log (1-\hat{y})\} + \lambda \left\|W\right\|_2,
\end{equation}
where $\Delta^+$ is the positive sample set and $\Delta^-$ is the negative sample set, $\left\|W\right\|_2$ is the L2 regularization to all the trainable parameters and $\lambda$ is the penalty weight.
Besides, we also apply dropout and early stopping to avoiding over-fitting.

\section{Experiments}
\subsection{Datasets}
We conduct experiments on a real-world online news dataset Adressa \cite{adressa}\footnote{http://reclab.idi.ntnu.no/dataset/}, which is a click log data set with approximately 20 million page visits from a Norwegian news portal as well as a sub-sample with 2.7 million clicks. Adressa is published with the collaboration of Norwegian University of Science and Technology (NTNU) and Adressavisen (local newspaper in Trondheim, Norway) as a part of RecTech project on recommendation technology, it is one of the most comprehensive open datasets for training and evaluating news recommender systems. 

The datasets are event-based including anonymized users with their clicked news logs. In addition to the click logs, the data set contains some contextual information about the users such as geographical location, active time (time spent reading an article), and session boundaries etc. We use the two  light versions, named Adressa-1week, which collects news click logs as long as 1 week (from 1 January to 7 January 2017), and  Adressa-10week, which collects 10 weeks (from 1 January to 31 March 2017) dataset.
Following DAN \cite{DAN}, for each event, we just select the (sessionStart, sessionStop)\footnote{sessionStart and sessionStop determine the session boundaries.}, user id, news id, time-stamp, the title and profile of news  for building our datasets.

Specifically, we first sort the news in chronological order. For the Adressa-1week dataset, we split the data as: the first 5 days' history data for graph construction
, the 6-th day's  for generating training pairs $<$u, d$>$, 20\% of the last day's for validation and the left 80\% for testing. Note that during testing, we reconstruct the graph with the previous 6 days' history data.
Similarly, for the Adressa-10week dataset, in training period, we use the previous 50 days' data for graph construction, the following 10 days' for generating  training pairs, 20\% of  the left 10 days' for validation  and 80\% for testing. 

To reduce the noise of textual data, we preprocess the data as follows. We remove the stopwords of the titles  and filter out meaningless entities and entity types \footnote{8 types of entities including site, author, language, adressa-importance, kundeservice-access, kundeservice-importance, adressa-access and pageclass are filtered out. And the remain 11 types of entities are: concept, sentiment, entity, classification, category, adressa-tag, person, location, company, taxonomy and acronym.} in the news profile.  
The statistics of our final datasets are shown in Table \ref{tab:tab1}.

\begin{table}[t]
\centering
\caption{Statistics of the dataset.}\label{tab:tab1}
\begin{tabular}{|l|c|c|}
\hline
Number & Adressa-1week & Adressa-10week \\  
\hline        
  \#users  & 537,627  & 590,673 \\
  \#news  & 14,732  & 49,994 \\
  \#events  & 2,527,571  & 23,168,411 \\
  \#vocabulary  & 116,603  & 279,214 \\
  \#entity-type  & 11  & 11 \\
  \#average words per title  & 4.03  & 4.10 \\
  \#average entity per news  & 22.11 & 21.29 \\
\hline
\end{tabular}
\end{table}

\subsection{Parameter Setting}
 We implement our model based on Tensorflow. The hyper-parameter settings are determined by optimizing $AUC$ on validation set. They are set as follows. The dimension of both word embeddings and entity type embeddings is set as $k_1=k_2=50$, and  the dimension of news embeddings, user embeddings and topic embeddings are set as $D=128$. The parameter configurations of parallel CNNs are set followed by DAN \cite{DAN}. The selected number of latest clicked news is set as the same ($l$=10) as  DAN \cite{DAN}. For LDA, the number of topics is set as $\mathcal{K}=20$. In GNN, the fixed number of sampled neighboring users, neighboring news documents are set as $L_u=10$ and $L_d=30$.
 
The embeddings are randomly initialized using a Gaussian distribution with a mean of 0 and a standard deviation of 0.1. And the parameters are optimized by Adam \cite{adam} algorithm with learning rate  0.0003. L2 penalty is set to 0.005 and the dropout rate is set to 0.5. We follow previous works \cite{DKN,DAN}, and use the same parameter settings for the baseline models. 

\subsection{Baselines}
We use the following state-of-the-art methods as baselines in our experiments:
\begin{itemize}
\item DMF \cite{DMF} is a deep matrix factorization model which uses multiple non-linear layers to process raw rating vectors of users and items. They ignore the news contents and take the implicit feedback as its input.
\item DeepWide \cite{DeepWide} is a deep learning
based model that combines the the linear model (Wide) and feed-forward neural network(Deep) to model low- and high-level feature interactions simultaneously. In this paper, we use the concatenation of news title and profile embeddings as features.
\item DeepFM \cite{DeepFM} is also a general deep model for recommendation, which combines a component of factorization machines and a component of deep neural networks that share the input to model low- and high-level feature interactions. We use the same input as in DeepWide for DeepFM.
\item DKN \cite{DKN} is a deep content based recommendation framework, which fuses semantic-level and knowledge-level representations of news by a multi-channel CNN. In this paper, following DAN \cite{DAN}, we model news title as semantic-level representations and profile as knowledge-level representations.
\item DAN \cite{DAN} is a deep attention based neural network for news recommendation, which improves DKN \cite{DKN}  by considering the user’s click  sequence information.
\end{itemize}

All the baseline models  are based on deep neural networks. DMF is a collaborative filtering based model, while the others are all content based.

\subsection{Experimental Results}
\subsubsection{Comparisons of Different Models}
In this subsection, we conduct experiments to compare our model with the state-of-the-art baseline models on two datasets, and report the results in Table \ref{tab:tab2} in terms of $AUC$ and $F1$ metrics. 
\begin{table}[t]
\centering
\caption{Comparison of Different Models}\label{tab:tab2}
\begin{tabular}{|l|c|c|c|c|}
\hline
\multirow{2}{*}{Model} &
\multicolumn{2}{c|}{Adressa-1week}&\multicolumn{2}{c|}{Adressa-10week}\\
\cline{2-5}
 & AUC(\%) & F1(\%) & AUC(\%) & F1(\%)\\
\hline        
  DMF  & 55.66  & 56.46 & 53.20  & 54.15\\
  DeepWide  & 68.25  & 69.32 & 73.28  & 69.52\\
  DeepFM  & 69.09  & 61.48 & 74.04  & 65.82\\
  DKN  & 75.57  & 76.11 & 74.32  & 72.29\\
  DAN  & 75.93  & 74.01 & 76.76  & 71.65\\
  GNewsRec & \textbf{81.16}  & \textbf{82.85} & \textbf{78.62}  & \textbf{81.01}\\
\hline
\end{tabular}
\end{table}

As we can see from Table \ref{tab:tab2}, our model consistently  improves all the baselines on both datasets by more than  10.67\% on F1 and  2.37\% on AUC. We attribute the significant superiority of our model to its three advantages: (1) Our model constructs a heterogeneous user-news-topic graph and learns  better  user and news embeddings with high-order information encoded by GNN. (2) Our model  considers not only the long-term user interest but also the short-term interest. (3) The topic information incorporated in the heterogeneous graph can help better reflect a user's interest and alleviate the sparsity issue of user-item interactions.  The news items with few user clicks can still aggregate neighboring information through the topics.

We also find that all content-based models achieve better performance than the  CF-based model  DMF. This is because CF-based methods cannot work well in news recommendation due to cold-start problem. Our model as a hybrid model can combine the advantages of content-based models and CF-based model. In addition,  new arriving documents without user clicks  can also be connected to the existing graph via topics, and update their embeddings through GNN. Thus, our model can achieve better performance. 

\subsubsection{Comparisons of GNewsRec Variants}
\begin{table}[t]
\centering
\caption{Comparison  of GNewsRec variants.}\label{tab:tab3}

\begin{tabular}{|l|c|c|c|c|}
\hline
\multirow{2}{*}{Model} &
\multicolumn{2}{c|}{Adressa-1week}&\multicolumn{2}{c|}{Adressa-10week}\\
\cline{2-5}
 & AUC(\%) & F1(\%) & AUC(\%) & F1(\%)\\
\hline        
    GNewsRec without GNN  & 75.93  & 74.01 & 76.76  & 71.65\\
    GNewsRec without short-term interest  & 79.00  & 80.53 & 77.03  & 80.21\\
   GNewsRec without topic  & 79.27  & 80.73 & 77.21  & 80.32\\
   GNewsRec  &\textbf{81.16}  & \textbf{82.85} & \textbf{78.62}  & \textbf{81.01} \\
\hline
\end{tabular}
\end{table}
Further, we compare among the variants of GNewsRec to demonstrate the efficacy of the design of our model with respect to the following  aspects:  GNN for learning user and news embeddings with high-order structure information encoded, combining of long-term and short-term user interests, and the incorporation of topic information.  The results are shown in Table \ref{tab:tab3}. 

As we can see from Table \ref{tab:tab3},  there is a great decline in performance when we remove the GNN module for modeling long-term user interest and news, which encodes high-order relationships on the graph. This demonstrates the superiority of our model by constructing a heterogeneous graph and applying GNN to  propagate the embeddings over the graph.

Removing short-term interest modeling module will decrease the performance by around 2\%  in terms of both AUC and F1. It demonstrates that considering both long-term and short-term user interests is necessary.  

Compared to the variant model without topic information, GNewsRec achieves significant improvements on both metrics. This is because that the topic information can alleviate the user-item sparsity issue as well as the cold-start problem. New documents with few user clicks can still aggregate neighboring information through topics. GNewsRec without topic performs slightly better than GNewsRec without short-term interest modeling, which shows that considering short-term interest  is important.

\subsubsection{Parameter Sensitivity}

In this section, we mainly explore the impact of different parameters of GNewsRec. We study   the impact of different number of GNN layers, and the effect of different dimension of news, user and topic embedding $D$ (which are set as the same).

\begin{figure}[t]
\centering 
\subfigure{
\label{AUC.eps}
\includegraphics[width=0.45\textwidth]{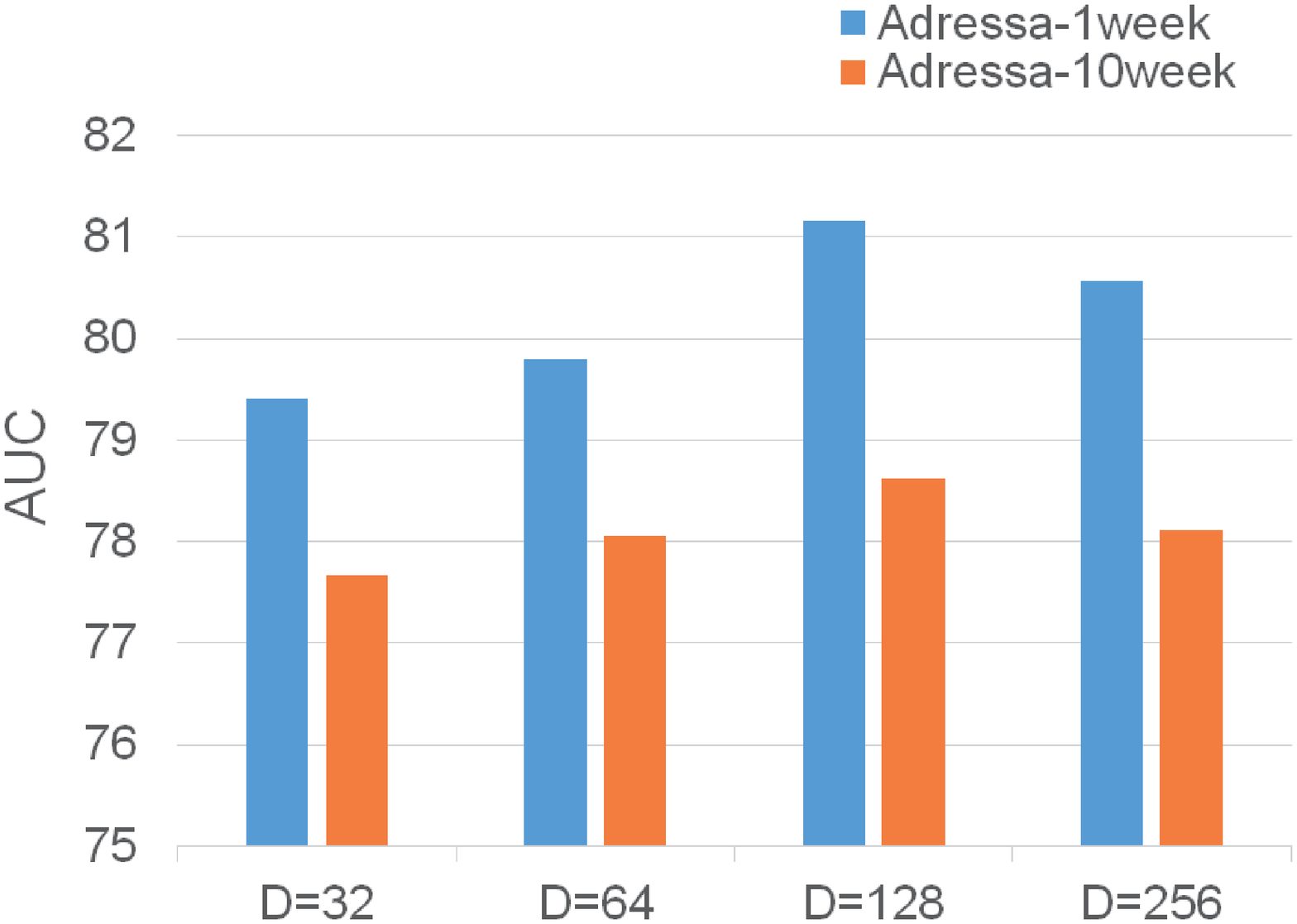}}
\subfigure{
\label{F1.eps}
\includegraphics[width=0.45\textwidth]{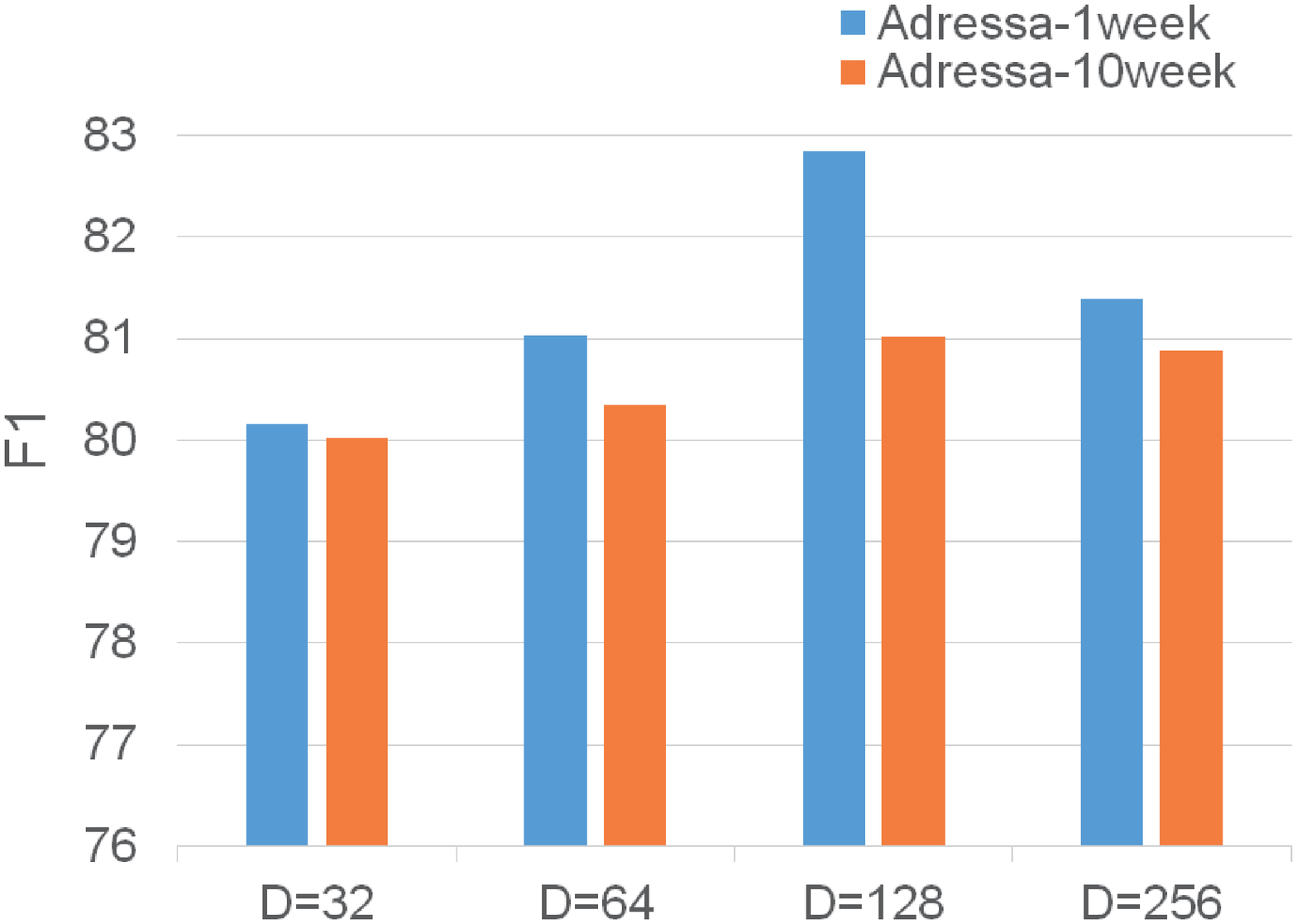}}
\caption{Dimension sensitivity of news embedding $D$}
\label{fig 3}
\end{figure}

\begin{table}[t]
\centering
\caption{Impact of different GNN layers of GNewsRec.}\label{tab:tab4}
\begin{tabular}{|c|c|c|c|c|}
\hline
\multirow{2}{*}{Model} &
\multicolumn{2}{c|}{Adressa-1week}&\multicolumn{2}{c|}{Adressa-10week}\\
\cline{2-5}
 & AUC(\%) & F1(\%) & AUC(\%) & F1(\%)\\
\hline        
 GNewsRec-1 layer  & 75.24  & 72.17 & 76.17  & 71.92\\
 GNewsRec-2 layers  &\textbf{81.16}  & \textbf{82.85} & \textbf{78.62}  & \textbf{81.01}\\
 GNewsRec-3 layers  & 78.94  & 80.36 & 77.92  & 80.11\\
\hline
\end{tabular}
\end{table}

We vary the number of GNN layers  from 1 to 3. From Table \ref{tab:tab4}, we can find that GNewsRec with 2-layer GNN performs best. This is because 1-layer GNN can't capture the higher-order relationships between users and news. Nevertheless, 3-layer GNN may bring massive noise to the model.  Higher layers with too long relation-chains make little sense when inferring inter-node similarities \cite{DKN}. Thus, we choose 2-layer GNN in our model GNewsRec.

We select the dimensions of $D$ in set $\{32, 64, 128, 256\}$. Figure \ref{fig 3} gives the convinced results, which are (1) Our model achieves the best performance at $D = 128$ setting, indicating than such dimension setting best express the semantic information of news, user and topic space. (2) The performance of our model  first increases with the growth of $D$ and then drops as $d$ further increases. This is because that too low dimension has insufficient capability of capturing the necessary information, and too large dimension introduces unnecessary noise and reduces generalization ability.

\section{Conclusion}
In this paper, we propose a novel graph neural news recommendation model GNewsRec  with long-term and short-term interest modeling. Our model constructs a heterogeneous user-news-topic graph to model user-item interactions, which alleviate the sparsity of user-item interactions. Then it applies  graph  convolutional networks to learn user and news embeddings with high-order information encoded by propagating embeddings over the graph. The learned user embeddings with complete historic user clicks are supposed to encode a user's long-term interest. We also model a user's short-term interest using recent user reading history with an attention based LSTM model. We combine both long-term and short-term interests for user modeling, which are then compared to the candidate news representation for prediction. Experimental results on a real-world dataset show that our model significantly outperforms state-of-the-art methods on  news recommendation.

\section{Acknowledgements}
This work is supported by the National Natural Science Foundation of China (No. 61806020, 61772082, 61972047, 61702296), the National Key Research and Development Program of China (2017YFB0803304), the Beijing Municipal Natural Science Foundation (4182043), the CCF-Tencent Open Fund, and the Fundamental Research Funds for the Central Universities.
\section*{References}

\bibliography{mybibfile}

\end{document}